\documentclass[12pt]{article}

\usepackage{ifthen}
\newboolean{shortdraft}
\newboolean{qcircuit}

\ifthenelse{\boolean{shortdraft}}{
\usepackage[paperheight=8in, paperwidth=6.5in, margin=0.2in]{geometry}
}{}

\usepackage{amsmath}
\usepackage{amsfonts}
\usepackage{amsthm}
\usepackage{amssymb}
\usepackage{url}
\usepackage{hyperref}

\ifthenelse{\boolean{qcircuit}}{
	\input{Qcircuit}
	\xyoption{all}
}{
\newcommand{\ket}[1]{\left | \, #1 \right \rangle}

}

\input{xy}
\xyoption{all}

\ifthenelse{\boolean{shortdraft}}{
	\renewcommand{\cite}[1]{[??]}
}{
	\bibliographystyle{halphads}
}

\title{Quaternionic quantum mechanics allows non-local boxes}
\author{Matthew McKague}

\begin{document}
\maketitle
\begin{abstract}
We consider non-local properties of quanternionic quantum mechanics, in which the complex numbers are replaced by the quaternions as the underlying algebra.  Specifically, we show that it is possible to construct a non-local box.  This allows one to rule out quaternionic quantum mechanics using assumptions about communication complexity or information causality.  
\end{abstract}

\section{Introduction}

We wish to shed some light on the structure of quantum mechanics by considering a question which is often on the minds of those just learning the theory:  Why a \emph{complex} Hilbert space?  We may justify this choice in a variety of ways (see for example, Scott Aaronson's discussion in \cite{Aaronson::PHYS771-Lecture}), citing algebraic closure, etc. but often one finds no statement about how the world would be \emph{different} if something else was chosen.  

What would happen if we replace the complex numbers with something else, leaving the remaining structure intact?  What could we reasonably replace them with?  Adler \cite{Adler:1995:Quaternionic-Qu} argues from a list of properties of probability amplitudes that the only possibilities are the reals, complex numbers, quaternions and octonions.  We will omit the octonions since they are non-associative.  Thus we will consider two alternatives to the complex numbers:  the reals and quaternions.
  
Recently it has been shown \cite{Matthew-McKague:2007:Simulating-Quan} \cite{McKague:2009:Simulating-Quan} that every experiment described in the usual quantum formalism admits another description within real quantum mechanics that predicts the same outcome statistics.  Moreover, the description within real quantum mechanics respects the same division into subsystems as the original description.  Thus the non-local effects of complex quantum mechanics are duplicated in real quantum mechanics.

In some sense, then, there is no difference between real and complex quantum mechanics; they both describe the same set of experiments and generate the same predictions for each of these experiments.  However, there are good reasons for preferring complex quantum mechanics.  For example, the translation from the complex description to the real description given in \cite{McKague:2009:Simulating-Quan} allows a larger set of operations in the real description than in the complex description, including operations corresponding to complex conjugation, and global phase measurements.  Since these operations do not appear to be physically realizable the complex formalism more closely resembles the experimental reality\footnote{Thanks to Bill Wootters for pointing this out.}.  Put another way, we may need some superselection rules in order to describe particular physical systems using the real formalism.

In the remainder of this paper we consider quantum mechanics over the quaternions.  A simulation, similar to the real simulation of complex quantum mechanics, developed by Fernandez et al. \cite{Fernandez:2003:Quaternionic-Co} shows that, for local systems at least, complex quantum mechanics can simulate quaternionic quantum mechanics.  But what about arbitrary systems with many subsystems?  We develop a construction for a non-local box (defined below) within quaternionic quantum mechanics.  Thus if we consider two or more subsystems, quaternionic quantum mechanics is distinguishable from complex quantum mechanics and exhibits much stronger non-local effects.

\section{Quaternionic quantum mechanics and the tensor product problem}

\subsection{Quaternions}

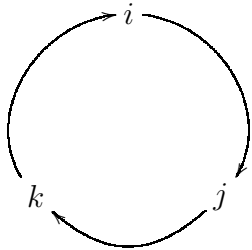
\begin{figure}
\[ \xymatrix {
 & i \ar@(r,ur)[ddr]& \\
  & & \\
 k \ar@(ul,l)[uur] & & \ar@(dl, dr)[ll]j \\}
\]
\caption{Multiplication in the quaternion group}
\label{fig:quatmult}
\end{figure}

The Quaternions ($\mathbb{H}$), first described by William Hamilton \cite{Hamilton:1844:On-quaternions-}, are a division ring formed by adjoining new elements, $i$, $j$, and $k$ to the real numbers $\mathbb{R}$.  Thus a quaternion looks like
\begin{equation}
q = a + ib + jc + kd.
\end{equation}
The new elements have the properties
\begin{equation}
i^{2} = j^{2} = k^{2} = ijk = -1.
\end{equation}
The multiplication of the elements $i, j, k$ is summarized in figure~\ref{fig:quatmult}.  When multiplying two elements along an arrow (eg. $ij$) the result is the third element in the cycle.  When multiplying backwards along an arrow (eg. $ji$) a $-1$ factor is added.  So $ij = k$ and $ji = -k$.  The elements $1$ and $-1$, of course, commute with the other elements.

Just as complex numbers have a real and imaginary part, quaternions have a scalar and vector part.  The scalar part is the part which lies on the real axis.  We denote it by $Re(q)$.  The vector part, also called the pure imaginary part, is everything else, and in general is a vector in $\mathbb{R}^{3}$.  We denote it by $Im(q)$.  The scalar and vector parts of $q$, defined above, are $R(q) = a$ and $Im(q) = ib + jc + kd$.

Like in the complex numbers, we may define  the \emph{conjugate} of a quaternion, which multiplies each of the non-real parts by -1.  Thus the conjugate of $q$ is
\begin{equation}
q^{*} := a - ib - jc - kd.
\end{equation}
The norm on the quaternions is the same as in the complex numbers, i.e.
\begin{equation}
||q|| = \sqrt{q q^{*}}
\end{equation}

The most important difference between the complex numbers and the quaternions is that the quaternions do not form a commutative algebra.  This property will be the basis for the rest of our discussion.

\subsection{Quaternionic quantum mechanics}
Quaternionic quantum mechanics is formed, roughly speaking, by replacing every complex number in the usual quantum mechanics by a quaternion.  Thus states are vectors over the quaternions, so amplitudes are now quaternions instead of complex numbers.  The usual norm-squared rule applies for deriving outcome probabilities, and discrete time evolution is described by unitary matrices $U$ over the quaternions with the usual property $U U^{\dagger} = I$.  Now the Hermitian conjugation ${(\cdot)}^{\dagger}$ is the matrix transpose, followed by quaternionic conjugation.  

Although many more aspects of quantum mechanics, such as continuous time evolution, may be considered, these few properties will suffice for our discussion.  For a comprehensive treatment of quanternionic quantum mechanics, see Stephen Adler's book \cite{Adler:1995:Quaternionic-Qu}.

\subsection{The tensor product problem}

The non-commutative nature of the quaternions introduces many new properties into quaternionic quantum mechanics.  The one we are most interested in here is the nature of multi-partite systems.

Consider a bipartite system in the state $\frac{1}{\sqrt{2}}\left(\ket{00} + \ket{11} \right)$.  Define the unitary matrices $R_{i}$ and $R_{j}$ as
\begin{equation}
R_{i} = \left(
	\begin{matrix}
	1 & 0 \\ 0 & i \\
	\end{matrix}
\right)
\end{equation}
\begin{equation}
R_{j} = \left(
	\begin{matrix}
	1 & 0 \\ 0 & j \\
	\end{matrix}
\right)
\end{equation}
We consider different ways that we may apply these matrices.  First we apply $R_{i}$ to the first subsystem, obtaining  $\frac{1}{\sqrt{2}}\left(\ket{00} + i\ket{11} \right)$.  Next we apply $R_{j}$ to the second subsytem, obtaining  
\begin{equation}\label{eq:stateij}
\frac{1}{\sqrt{2}}\left(\ket{00} - k\ket{11} \right).
\end{equation}  
Now consider the same operations, but applied in the reverse order.   We apply $R_{j}$ to the second subsystem, obtaining $\frac{1}{\sqrt{2}}\left(\ket{00} + j\ket{11} \right)$, followed by $R_{i}$ applied to the first subsystem, obtaining
\begin{equation}\label{eq:stateji}
\frac{1}{\sqrt{2}}\left(\ket{00} + k\ket{11} \right).
\end{equation}
Here we see the non-commutativity of $\mathbb{H}$ in action.  The two states in equations~\ref{eq:stateij} and~\ref{eq:stateji} are orthogonal, but all we have changed is the time-ordering of two local operations on separate subsystems.

The above problem may be stated as follows:  $R_{i} \otimes I$ and $I \otimes R_{j}$ do not commute.  This extends to the tensor product problem:  How do we define $R_{i} \otimes R_{j}$?  Evidently the evolution of subsystems cannot be considered without considering the context of the system as a whole.  Adler \cite{Adler:1995:Quaternionic-Qu} considers the same problem in the context of continuous evolution:

\begin{quote}
We conclude, then, that in quaternionic quantum mechanics, a sum of $N \geq 2$ 
one-body Hamiltonians gives a many-body Hamiltonian that does not describe 
N independent particles; the particle motions are coupled through the 
noncommutativity of the quaternion algebra.  (Adler \cite{Adler:1995:Quaternionic-Qu}, p. 245)
\end{quote}

What does this mean for locality?  Is there such thing as a local transformation?  Is it possible for Alice and Bob to actually perform the operations $R_{i} \otimes I$ and $I \otimes R_{j}$?  The formalism does not answer this question.  However, we may address this problem in another way.  If Alice and Bob can locally perform these operations, then they can implement a non-local box.

\section{Non-local boxes}
The non-local box, first defined by Popescu and Rohrlich in \cite{Popescu:1994:Quantum-nonloca}, is an imaginary device which produces non-local correlations between data in the following way:  Two distant parties, Alice and Bob, each have half of the box.  They have one bit of input each, $a$ and $b$, and input their bit into their half of the box.  Each half of the box produces one bit of output, $x$ and $y$, obeying the property
\begin{equation}
x \oplus y = ab.
\end{equation}
The content of the famous CHSH inequality \cite{Clauser:1969:Proposed-Experi} is that this condition cannot be satisfied by a non-signalling classical local hidden variable theory with probability better than $0.75$ when $a$ and $b$ are chosen uniformly at random.  Quantumly, we can do better, but are bounded above by $\cos^{2} \pi / 8 \approx 0.85$, the Cirel'son bound \cite{Cirelson:1980:Quantum-general}.

Now we consider the case of quaternionic quantum mechanics.  Evidently it has stronger non-local behaviour than complex quantum mechanics, but how strong?  Clearly we can at least achieve the Cirel'son bound since any strategy in complex quantum mechanics also exists in quaternionic quantum mechanics, but can we do better?  The answer is that we simulate the non-local box perfectly.

Consider the two parties, Alice and Bob, as before.  Before receiving their inputs they synchronize clocks and choose times $t_{1} < t_{2}< t_{3}< t_{4} < t_{5}$ such that $t_{1}$ is after they receive their inputs and $t_{5}$ is before they require the outputs (we may arrange it so that the time elapsed is too short to allow signalling by moving Alice and Bob far enough away from each other.)  They also share the state $\frac{1}{\sqrt{2}}\left(\ket{00} + k\ket{11} \right)$.

Alice does the following:

\begin{enumerate}
	\item Receive input $a$.
	\item If $a = 0$ then apply operation $R_{i}$ at time $t_{1}$.
	\item If $a = 1$ then apply operation $R_{i}$ at time $t_{3}$.
	\item At time $t_{5}$ measure in the basis $\ket{+} / \ket{-}$ and output the result as $x$.
\end{enumerate}

Meanwhile, Bob does the following:

\begin{enumerate}
	\item Receive input $b$.
	\item If $b = 0$ then apply operation $R_{j}$ at time $t_{4}$.
	\item If $b = 1$ then apply operation $R_{j}$ at time $t_{2}$.
	\item At time $t_{5}$ measure in the basis $\ket{+} / \ket{-}$ and output the result as $y$.
\end{enumerate}

Roughly what is happening here is that Alice applies her operation before Bob in all cases except when both of their inputs are 1.  Alice and Bob then detect this event using local measurements that are correlated except when Bob goes first, in which case they are anti-correlated.

Consider the outputs that Alice and Bob generate.  First note that the final state before measuring will be $\ket{00} \pm \ket{11}$.  If Alice and Bob both measure in the basis $\ket{+} / \ket{-}$ their outcomes will be the same if the relative phase was $+$ and opposite if the relative phase was $-$.

Suppose that Bob's input is $0$.  He will wait until $t_{4}$ before applying $R_{j}$.  Regardless of her input, Alice will apply $R_{i}$ before Bob applies $R_{j}$.  Thus the combined effect on $\ket{\psi}$ is a relative phase change of $-k$, taking the state to $\ket{\phi_{+}}$.  Then Alice and Bob's measurements will agree and $x \oplus y = 0 = ab$.

Meanwhile, if Bob's input is $1$ the situation changes.  If Alice receive the input 0 then she applies $R_{i}$ at time $t_{1}$ and Bob applies $R_{j}$ at time $t_{2}$ and the situation is the same as above.  However, if Alice receives the input 1 then she applies $R_{i}$ at time $t_{3}$, \emph{after} Bob applies $R_{j}$ at time $t_{2}$.  In this case the effect on $\ket{\psi}$ is a relative phase change of $k$, taking the state to $\ket{\phi_{-}}$.  In this case Alice and Bob's measurements will disagree and $x \oplus y = 1 = ab$.

\section{Discussion}

We now briefly consider \emph{communication complexity}.  Suppose two parties, Alice and Bob, receive two inputs, $a$ and $b$.  They wish to compute the value of some funciton $f(a,b)$.  How much communication is necessary between Alice and Bob? (for simplicity, we suppose that Alice receives the final answer.)  Here we are interested in boolean functions, whose output is a single classical bit.  It has been shown by van Dam \cite{Dam:2005:Implausible-Con} that the communication complexity of all boolean functions is trivial if non-local boxes are available.  This means that Bob needs to send only one bit of information to Alice, and Alice does not have to send anything to Bob.  However, there exist boolean functions, such as the inner product between two strings, for which the communication complexity in either a classical or quantum setting is maximal (i.e. the optimal strategy is for Bob to transmit his entire input to Alice) \cite{Brassard:2006:Limit-on-Nonloc}.  Coupled with the current result we find that within quaternionic quantum mechanics the communication complexity of all boolean functions is trivial.

Later Brassard et al. \cite{Brassard:2006:Limit-on-Nonloc} turn van Dam's result around, saying that if there is a non-trivial bound on the communication complexity of boolean functions, then non-local boxes do not exist.  They also made this result robust by introducing a notion of probabilistic communication complexity and showing that if a non-local box can be approximated with probability better than $\approx .906$ then every boolean function has trivial probabilistic communication complexity.  Linden et al \cite{linden:180502} finally showed that for a particular boolean function (AND of two 2-bit strings) if a non-local box can be approximated with probability better than $\cos^{2} \pi/8$ (the quantum upper bound) then no communication is required and the function can be approximated better than the classical (and quantum) bound of $0.75$.  Turning this result around, if the communication complexity of AND of 2-bit strings is non-trivial, then the non-local box cannot be approximated any better than what is achievable by quantum mechanics.  In particular, if there is a non-trivial bound on communication complexity then quaternionic quantum mechanics is not a viable theory.

Following this work, Pawlowski et al. \cite{Pawlowski:2009:A-new-physical-} developed the notion of \emph{information causality} which can be seen as a generalization of the notion of no-signalling.  Both classical and quantum theories obey information causality.  Pawlowski et al. were able to show that any physical theory which obeys information causality also obeys the Cirel'son bound \cite{Cirelson:1980:Quantum-general}.  Thus this is another way of excluding quaternionic quantum mechanics as a viable physical theory.

{\bf Acknowledgements}  This work is supported by NSERC, Ontario-MRI, OCE, QuantumWorks, MITACS, and the Government of Canada.  Thanks to William Wootters, Michele Mosca and Lana Sheridan for helpful discussions.

\ifthenelse{\boolean{shortdraft}}{}{
\bibliography{Global_Bibliography}
}
\end{document}